------------------------------------------------------------------
\def\ov{\over}
\def\d{\delta}
\def\p{\phi}
\def\h{\rho}
\def\h1{\rho_1}
\def\h3{\rho_3}
\def\h2{\rho_2}
\def\y1{y_1}
\def\y2{y_2}
\def\y3{y_3}

\def\h{\rho}
\def\h1{\rho_1}
\def\h3{\rho_3}
\def\h2{\rho_2}
\def\y1{y_1}
\def\y2{y_2}
\def\y3{y_3}

\def\dy{\delta y}

\def\d{\delta}
\def\h{\rho}
\def\h1{\rho_1}
\def\h3{\rho_3}
\def\h2{\rho_2}
\def\y1{y_1}
\def\y2{y_2}
\def\y3{y_3}

\def\d{\delta}

\def\dy{\delta y}
\def\dy{\delta y}

\documentstyle[ichep,12pt]{article}
\title{\normalsize INTERMITTENCY AND PARTICLE CORRELATIONS:
       THEORY}
\author{ Ina Sarcevic\\
Department of Physics \\
University of Arizona \\
Tucson, AZ 85721}

\begin{document}
\finalcopy

\maketitle

\abstract{
We review recent theoretical developments in understanding
the intermittency phenomenon observed
in high-energy leptonic, hadronic and nuclear
collisions.  In particular, we discuss
self-similar cascading and QCD parton showers,
models with phase transitions,
and the three-dimensional statistical field theory
of multiparticle density
fluctuations.}
\vskip-1pc
\onehead{INTRODUCTION}

The first unusually large local fluctuations in
rapidity distribution were observed
in a high multiplicity cosmic ray event \cite{one}, followed by the
famous NA22 ``spike'' event \cite{two}.
The analysis proposed to study these fluctuations by measuring the factorial
moments, which act as filters for the spike events, showed that
these spectacular events were not the result of
statistical fluctuations.
The observation of a power-law behavior of the factorial moments, i.e.
$F_p \sim \delta y^{-\nu_p}$,
in a
sufficiently large range of rapidity scales,
$\delta y$,
was claimed to be
a signal
of a dynamical
``intermittent'' behavior, in analogy with the onset of turbulence
in hydrodynamics \cite{three}.  In particle physics, this would correspond
either to a self-similar cascade process or to the behavior of
a statistical
system near a critical point.
The simplest example of the self-similar cascade is the chain
decay of hadronic ``clusters",
the initial heavy-mass ``cluster" decaying into smaller clusters,
which in turn decay into still smaller clusters.
This
leads to a power-law behavior for the normalized multiplicity
moments, with exponents related to the (multi-)fractal
dimension \cite{4}.
On the other hand, if a system undergoing a second-order
phase transition is close to its critical point, the
correlation functions exhibit power-law singularities.
This
implies that scaling laws and intermittency
exponents are related to anomalous dimensions
representing a simple (mono-) fractal \cite{5}.
The attractive idea of using intermittency (i.e.
multiparticle fluctuations present for a large range of scales)
to study the
fractal structure
of high-energy collisions
has inspired extensive experimental and theoretical work \cite{6}.

\onehead{ EXPERIMENTAL RESULTS}

In the past few years, several experimental groups have
investigated intermittency signals by
measuring factorial moments defined as \cite {three}
$$F_p (\d) =
                     {<{n (n -1) \ldots (n -p+1)>} \over {<n>}^p}=$$
$$=  {1 \over M (\d)^p } \sum_{m=1}^M
                     \int_{\Omega_m} \prod_i d^3x_i  \;\;
                     {\rho_p(\vec{x}_1 \ldots \vec{x}_p )
\over (\bar\rho_m)^p }
\eqno(1)$$
where $n$ is the number of particles in a bin $m$, $M$ is the
total number of bins, $\d$ is the
phase space region and
$\rho_p$ is p-particle density correlation function.
Initially, the analysis was done in rapidity (i.e. $\d\equiv \dy$ and
$\d y=Y/M$)
and factorial moments
were found to increase
with decreasing bin size \cite{6}.  Many experiments claimed
to observe
intermittency,
i.e. $F_p\sim \delta y^{-\nu_p}$, even though all the data
clearly showed
a tendency to level off at small $\delta y$.
It has been shown
that all one-dimensional hadronic
data can be described by exponential
two-particle correlations
and the linked-pair ansatz for higher-order correlations,
without invoking
any singular behavior for the correlations \cite{7}.  Similar
conclusions were reached for one-dimensional leptonic and
nuclear data \cite{8}.
Recently, Ochs has pointed out the importance of
performing the experimental
analysis in three dimensions (rapidity, $p_T$ and azimuthal angle),
and has indicated how projection onto two or one dimension can reduce
or destroy the ``intermittency'' signal \cite{9}.  In the past year,
multidimensional analyses have been performed for $e^+e^-$ collisions
(by
the CELLO,
DELPHI, ALEPH and OPAL Collaborations), for hadronic collisions
(by the NA22 and UA1 Collaborations) and
for nuclear collisions
(by the KLM, EMU01 and NA35 Collaborations) \cite{10}.
As predicted by Ochs,
all experiments observe a stronger intermittency signal in higher
dimension.
In $e^+e^-$ collisions, all the multidimensional, as well as one dimensional,
data were found to be
consistent with the existing parton shower Monte Carlos, indicating that
the observed increase of the factorial moments with decreasing
phase space size
is the consequence of the parton cascade.
In the case of hadronic collisions, none of the existing Monte Carlos
can account for the observed effect and it seems likely
that it is some combination of a
perturbative parton
cascade with soft-type interactions.  Experimental analysis of
the
semi-hard (i.e. ``minijet'') events could help unravel
this
problem.
In high-energy
heavy-ion collisions,
there is
no theory which
explains the observed rise of the moments and all the existing
Monte Carlo programs for heavy-ion collisions fail to reproduce
the data.
It has been argued that
the unusually large fluctuations
signal
a phase transition from quark-gluon plasma to hadronic matter.
So far, no conclusive prediction for the
signature of quark matter has been identified
and, as a consequence the study
of the unusually large density fluctuations
has attracted considerable attention, especially in as much as they
may reveal the
presently unknown dynamics of particle production.

\onehead{THEORETICAL MODELS}

\twohead{Cascade Models}

\par
The simplest multidimensional model that leads to intermittency is
the so-called $\alpha$-model \cite{three}.  In this model,
the particle density
in each sub-interval
is a product of random numbers with common distribution
so that the cascade is self-similar.  The factorial moments are
power-laws, while the one-dimensional projections show saturation
at the small $\delta y$ (i.e. the intermittency signal gets lost).
This model does not
describe the data well and should be used
only as a toy-model.  In $e^+e^-$ collisions, the dominant mechanism
for particle production is a QCD parton cascade, which
implicitly violates
scaling because of the
running coupling constant,
the
angular cutoff,
and the formation of hadronic resonances.  It has been
shown
that statistical models, such as those that employ
log-normal and
negative binomial distributions, fail to reproduce the
data in the
small region of phase space, confirming the importance of
QCD parton cascading in the underlying dynamics \cite{fos}.
In the case of high-energy
hadronic collisions, one can construct a simple
self-similar cascading model for multiparticle production.
At very high energies,
the phase space available for particle production
is large enough to allow a self-similar
cascade with many branches to develop.  Clearly, this new
mechanism for multiparticle production has some threshold energy.
For example, at $\sqrt s=20GeV$ the cascade has only a
few
branches, since the maximum rapidity available ($Y\equiv\ln  s$)
is only slightly
above
the resonance formation threshold.  In contrast,
at SSC energies,
$Y \geq 20$ and
a self-similar cascade with many branches can develop.
The threshold energy for this self-similar cascade
mechanism is
of the order of a few hundred GeV.  Since we expect that at these
energies the application of perturbative QCD
is well justified, this self-similar cascade should be
related
to low-$p_T$ jet production.
In support of this conjecture,
we mention the recent observations of ``minijets'',
which indicate that the
fraction of ``semi-hard'' events
responsible for low-$p_T$ jet production
increases very quickly with energy.  For example,
at $\sqrt s\sim 20-50GeV$
it is only a few percent, while at CERN Collider energies,
it
is about $15-17\%$.  At SSC energies, one expects that
most of the events will be ``semi-hard''.
In the simple
self-similar cascade model \cite{4},
a collision takes place
in several steps.  First, a ``heavy mass particle'' is created (this
could be a jet, for example).  Then, this particle
decays into two lighter particles of mass $m_1$, which by the
conservation
of energy is related to the
mass $m_0$
of the initial particle
($m_0^2=4(m_1^2 + p_1^2)$).
This pattern
continues until the initial
mass is reduced to the mass of the
resonance ($m_{\pi \pi}\sim 0.5GeV$).
Such a one-dimensional self-similar cascade model
is found to agree well
with the UA5 data on multiplicity moments
in different rapidity regions \cite{4}
and it would be very interesting to test
it at
Tevatron and SSC energies.
With a better understanding of nonperturbative
QCD, one might be
able to derive
the intermittency exponents, which in this case
resemble multifractals.  Deriving the
probability distribution by solving stochastic
evolution equations, in which quarks and gluons branch with
Altarelli-Parisi-type probabilities, has also been
shown to reproduce much of the data
\cite{12}.  A model that contains
two mechanisms has been proposed,
in which parameters of the negative binomial
distribution have a physical interpretation
\cite{13}.  Furthermore,
multifractality and phase transitions have
been studied in a two-component self-similar model,
that purports to identify the
signature of the formation of the
quark-gluon plasma
\cite{14}.

\twohead{Intermittency and Phase Transitions}

The study of intermittency in the 2D Ising Model provides
a simple illustration of the connection between the
critical
behavior or scale invariance of the underlying theory
and the corresponding
intermittency
exponents \cite{5}.
The intermittency exponents ($\nu_p$) of the block
spin moments
were found to be equal to
$D(p-1)$, where D is
related to the critical exponents of the Ising Model.  Bialas and
Hwa \cite{15} conjectured that data that follow
this behavior
indicate that the system has undergone
a phase transition from a quark-gluon plasma to a
hadronic gas.
By fitting the one-dimensional
factorial moments with straight lines (i.e. by
assuming that the moments
do not saturate at small $\delta y$) and then by determining
whether these
slopes follow monofractal (signal of quark-gluon plasma)
or multifractal
(signal of the cascading) behavior,
they claimed that the QGP has been observed in
S-Au data.
These
data were later reanalyzed and
somewhat different exponents were found,
$\nu_p=(p^{1.6}-1)/(p-1)$ \cite{9}.
Recently, the exact solution for the factorial moments in
the
1D Ising Model have been shown to display
intermittency and these
results have been compared to
high-energy nuclear data \cite{16}.
The universality of the factorial moments, namely that all the
moments, $F_p$, can be expresses as some power of $F_2$
($F_p \sim
{(F_2)}^{c_p}$), was found to be present in all high-energy
collisions \cite{17}.  The fundamental reason for this behavior
is still not understood.
Finally, there have been some
attempts to interpret hadronic data in terms of a
second-order phase transition by evoking a
two-component model (one component corresponds to
the Feynman-Wilson ``gas'' and the other to
the critical behavior) \cite{18}.

\twohead {Cumulant Expansion and Factorial Cumulants}

In order to examine the true higher-order correlations, the
trivial contributions from two-particle correlations need to
be subtracted.
The connection between the factorial
moments and the correlations can be seen in Eq. (1).
The $F_p$ can be expressed
in terms of the bin-averaged cumulant moments:
$$F_2 = 1+ K_2
$$
$$F_3 = 1+ 3K_2 +K_3 $$
$$F_4 = 1+ 6K_2 + 3 {(K_2)^2} + 4K_3 + K_4 \eqno(2)$$
where
$$K_p (\d) =
{1 \ov M(\d)^p} \sum_m \int_{\Omega_m} \prod_i d^3x_i
{}~~~{k_p (\vec{x}_1 \ldots \vec{x}_p ) }$$
and
$$k_2(1,2)= {\rho_2(\vec{x}_1,\vec{x}_2)\over {<\rho
(\vec{x}_1)><\rho(\vec{x}_2)>}}-1 \eqno(3)$$
$$k_3(1,2,3)={\rho_3(1,2,3)
\over {<\rho(1)><\rho(2)>
<\rho(3)>}}-$$
$$
-\sum_{perm}^{(3)} {\rho_2(1,2)\over
{<\rho(1)><\rho(2)>}}
+ 2 .$$
Clearly, if there are no true, dynamical correlations,
the cumulants $K_p$ vanish.

It has been found that
$K_2$ decreases from lighter to heavier projectiles
, especially in the case of Sulfur.
Furthermore, in hadronic collisions $K_3$ and $K_4$ are
non-negligible (for example, $K_3$
contributes up to $20\%$ to $F_3$ at small
$\dy$), while
in nucleus-nucleus collisions, at the same energy,
these cumulants are compatible with zero \cite{19}.
This implies that there are no statistically significant
correlations of order higher than two for heavy-ion
collisions (i.e.
the observed increase of the higher-order factorial
moments $F_p$ is entirely due to the
dynamical two-particle correlations).
This conclusion was found to hold even in a higher-dimensional
analysis \cite{20}.
\par
It is intuitively clear that rescattering of initially correlated
particles by downstream constituents should decorrelate those initial
correlations.
More quantitative calculations
are needed to
explain these phenomenological results, in particular
for the anticipated rapidity fluctuations at RHIC and
LHC energies, in order to see whether suppression of multiplicity
cumulants, and the attendant dominance of factorial moments by
two-particle cumulants, continues to hold.  Even if strong
space-time fluctuations should occur, of the sort
associated with the transition
to a quark-gluon plasma phase, the rapidity moments must obey the
identities of Eqs. (2).  In this case, however, we expect the higher
cumulant moments to suddenly increase, to reflect the presence
of the more violent bulk fluctuations that precede hadronization.

\twohead{Statistical Field Theory of Multiparticle Density
Fluctuations}

We have seen that particles
produced in high-energy heavy-ion collisions exhibit
only two-particle correlations, indicating that perhaps
higher-order correlations are washed out by rescattering
of the initially correlated particles.  Presently, there is
 no theory that describes this phenomena.
Recently, a three-dimensional
statistical field theory of density
fluctuations which has these features has been proposed \cite{21}.
This model was formulated in analogy with the Ginzburg-Landau
theory of superconductivity \cite{22}.  The large
number of particles produced in ultrarelativistic
heavy-ion collisions justifies the use if a
statistical theory of particle production.
The formal analogy with the statistical mechanics
of a one-dimensional ``gas'' was first pointed out by
Feynman and Wilson \cite{23} and was
later further developed by
Scalapino and Sugar \cite{24} and many others \cite{25}.
The idea is to build a
statistical theory of the macroscopic observables by
imagining that the microscopic degrees of freedom are
integrated out and represented in terms of a few
phenomenological parameters
and by postulating that
this theory
will
eventually be derived from a more
fundamental theory, such
as QCD.
\par
While in the G-L theory of superconductivity
the field (i.e. the order parameter) represents superconducting
pairs, in the particle production problem, the relevant
variable is the density fluctuation.
The ``field'' $\phi(\vec{x})$ is
a random variable which depends on the rapidity of the
particle and
its transverse momentum $p_t$ and it is
identified with the density fluctuation
(specifically,
$\p(\vec{x})={\rho(\vec{x})\over {<\rho(\vec{x})>}} - 1,$
so that $<\p>\equiv 0$).
\par
Even though particles
produced in high-energy collisions need not be in
thermal equilibrium, one can still introduce a
functional of the field $\phi$, $F[\phi]$, which plays
a role analogous to the free energy in equilibrium
statistical mechanics.  In principle one should be
able to derive this functional from the underlying
dynamics.
\par
We start by introducing our functional $F[\phi]$ in the
Ginzburg-Landau form
$$
F[\phi]=\int_0^Ydy\int_{p_t\leq p_{t,{\rm max}}}
\frac{d^2p_t}{P^2}~[a^2(\partial _y\phi )^2+$$
$$
+a^2(\nabla _{(p_t/P)}\phi )^2+M^2\phi ^2+V(\phi )]\ ,
$$
\noindent
where
$Y$ is the rapidity gap between projectile and target,
and $a$ and $M$ are phenomenological parameters
that depend on control parameters of the
considered reaction (such
as total energy and mass number(s)).
All physical quantities can be obtained
in terms of
ensemble averages
appropriately weighted by $F[\phi]$ with the
corresponding ``partition function''
$Z=\int {\cal D \phi} e^{-F[\phi]}$.
For example,
field correlations, which
from our definition of the field
are related to particle correlations, are given by

$$
<\phi(\vec{x_1})\phi(\vec{x_2})...\phi(\vec{x_p})> =
$$
$$={1\over Z}
\int {\cal D \phi} e^{-F[\phi]} \phi(\vec{x_1}) \phi(\vec{x_2})...
\phi(\vec{x_p}),
$$
where $\vec{x}\equiv (y,\vec{z}\equiv\vec{p_t}/P)$.

\par

In particular, using the definition of the field
one finds that the field
correlations correspond to the following
cumulant particle
correlations $k_p$:
$$<\phi(\vec{x_1})\phi(\vec{x_2})>\equiv k_2(1,2)$$
$$<\phi(\vec{x_1})\phi(\vec{x_2})\phi(\vec{x_3})>
\equiv k_3(1,2,3)\eqno(4)$$
$$<\phi(\vec{x_1})\phi(\vec{x_2})\phi(\vec{x_3})
\phi(\vec{x_4})>\equiv k_4(1,2,3,4)
$$
$$+ \sum_{perm}^{(3)}
k_2(1,2)k_2(3,4),$$
where the $k_p$'s are defined by Eqs. (3).

Clearly, {\it if} the interaction term in the functional $F[\phi]$
is not present (if
$V(\phi)\equiv 0$),
all the higher-order odd-power correlations vanish
and even-power correlations
can be expressed in terms of two-field correlations.
This implies that $k_p=0$ for $p\geq 3$, while
the two-particle correlations are given by
$$
\langle\phi(\vec{x_1})\phi(\vec{x_2})\rangle =
\frac{\gamma}{2\pi\xi}{\rm e}^{-\vert \vec{x_1}-
\vec{x_2}\vert/\xi}/\vert \vec{x_1}-\vec{x_2}\vert
$$
$$\langle\phi(y_1)\phi(y_2)\rangle
= \gamma {\rm e}^{-\vert y_1-y_2\vert/\xi}\ ,
$$
where $\gamma=1/4aM$ and
$\xi=a/M$ and the second equation applies for the
one-dimensional case considered below. Note that the
three-dimensional correlation function
has a singular, Yukawa-type form.
\par
In the three-dimensional
case, $K_2$ is obtained by numerical integration over
the appropriate phase space region.  The results are found
to be in good agreement with multidimensional data \cite{26}.
The three-dimensional cumulant obeys a power-law behavior, i.e.
$K_2(\delta)\sim 1/\delta$,
as observed experimentally, and all the two-dimensional and
one-dimensional projections saturate in the small
bins.  The results for one-dimensional field theory
can be found in Ref. 21.
Both coefficients, ``mass'' $M$
and ``kinetic coefficient'' $a$,
are found to increase with
the complexity of the system.  The value of
the correlation length $\xi$
usually determines
how far the system is from the critical point.
When $\xi\rightarrow \infty$ (or $M\rightarrow 0$), the
system goes through the phase transition.  The
fitted values for $\xi$ ($\xi\sim O(1)$)
do not indicate critical behavior for
the system at present energies.
Approaching a critical
point or, more generally,
a phase transition will presumably change
the behavior of the two-particle as well as
the multi-particle correlations.
Therefore, the
appealing possibility that we may study the phase transition from
hadronic matter to a quark-gluon plasma and provide
further constraints on theoretical models
by measuring
three-dimensional density fluctuations at higher energies (e.g.
at RHIC and LHC)
certainly deserves
further investigations.

\onehead{CONCLUSIONS}

Recent measurements of
intermittency and particle correlations
in leptonic, hadronic and heavy-ion
collisions show
the presence of unusually large
nonstatistical fluctuations in a range of
phase space intervals.
All $e^+ e^-$ data are found to be consistent
with the
Standard Parton Shower Monte Carlos,
confirming the
importance of
QCD as the underlying dynamics.  However,
there is
no Monte Carlo that describes the hadronic data, where
particle correlations (i.e. intermittency)
become even stronger at higher
energies.  An interesting possibility is that the
particle production at high energies is govern by the
self-similar cascading of the hadronic ``clusters'' or
low-$p_T$ jets, resulting in scaling behavior of the
multiplicity moments with multifractal exponents.
The phase space available at
the SSC and LHC is large enough
that one might be able to search for
fractal structure in
multiparticle production and thereby, gain
insight into the scale invariance of
``soft'' partonic cascades.
Since in
high-energy heavy-ion
collisions, strong collective effects have been observed,
the possibility of creating the quark-gluon plasma at RHIC and
LHC seems realistic.  However,
whether and how one can use intermittency/multiparticle
correlations to
detect this new phase of matter
is not clear and deserves further theoretical
study, in particular by connecting present ``models''
to the fundamental
theory of the strong interactions, QCD.

\end{document}